\documentclass[a4paper,12pt]{article}

\usepackage{citesort}
\usepackage{epsfig}

\makeatletter

\@addtoreset{equation}{section}
\makeatother
\def\lsim{\;\raise0.3ex\hbox{$<$\kern-0.75em\raise-1.1ex\hbox{$\sim$}}\;}
\def\gsim{\;\raise0.3ex\hbox{$>$\kern-0.75em\raise-1.1ex\hbox{$\sim$}}\;}

\def\wt{\widetilde}
\def\ben{\begin{enumerate}}  \def\een{\end{enumerate}}
\def\bit{\begin{itemize}}    \def\eit{\end{itemize}}
\def\beq{\begin{equation}}   \def\eeq{\end{equation}}
\def\ba{\begin{array}}       \def\ea{\end{array}}
\def\bea{\begin{eqnarray}}   \def\eea{\end{eqnarray}}
\def\nn{\nonumber}
\def\noi{\noindent}

\begin{document}

\begin{titlepage}
\renewcommand{\thefootnote}{\fnsymbol{footnote}}
\setcounter{footnote}{0}

\begin{flushright}
LPT Orsay 99-65 \\
\end{flushright}

\begin{center}
\vspace{3cm}
{\Large\bf Masses and Couplings of the Lightest \\ Higgs Bosons in the
(M+1)SSM} \\
\vspace{2cm}
{\bf Ulrich Ellwanger\footnote{email: ellwange@th.u-psud.fr} and Cyril
Hugonie\footnote{email: cyrilh@th.u-psud.fr}} \\
Laboratoire de Physique Th\'eorique\footnote{Unit\'e mixte de Recherche (UMR
8627)} \\
Universit\'e de Paris XI, F-91405 Orsay Cedex, France
\vspace{2cm}
\end{center}

\begin{abstract}

We study the upper limits on the mass of the lightest and second lightest CP
even Higgs bosons in the (M+1)SSM, the MSSM extended by a gauge singlet. The
dominant two loop contributions to the effective potential are included, which
reduce the Higgs masses by $\sim 10$~GeV. Since the coupling $R$ of the
lightest Higgs scalar to gauge bosons can be small, we study in detail the
relations between the masses and couplings of both lightest scalars. We present
upper bounds on the mass of a 'strongly' coupled Higgs ($R > 1/2$) as a
function of lower experimental limits on the mass of a 'weakly' coupled Higgs
($R < 1/2$). With the help of these results, the whole parameter space of the
model can be covered by Higgs boson searches.

\end{abstract}

\end{titlepage}

\renewcommand{\thefootnote}{\arabic{footnote}}
\setcounter{footnote}{0}

\section{Introduction} \label{sec:int}

Curiously enough, the most model independent prediction of supersymmetric
extensions of the Standard Model concerns a 'standard' particle: the mass $m_h$
of the (lightest CP even) Higgs boson. Within the minimal supersymmetric
extension of the Standard Model (MSSM) its mass is bounded, at tree level, by

\beq
m_h^2 \leq M_Z^2 \cos^2 \! 2\beta
\eeq

\noi where $\tan\!\beta = h_1/h_2$ ($H_1$ couples to up-type quarks in our
convention). It has been realized already some time ago that loop corrections 
weaken this upper bound \cite{r1}. These loop corrections depend on the top
quark Yukawa coupling $h_t$ and the soft susy breaking parameters as the stop
masses of $O(M_{susy})$. At the one loop level, given the present experimental
errors on the top mass $m_t$ and assuming $M_{susy} \lsim 1$~TeV, the upper
limit on $m_h$ is $\lsim 140$~GeV. Also two loop corrections to $m_h$ have been
considered in the MSSM \cite{r2,r3,r4}; these have the tendency to lower the
upper bound on $m_h$ by $\sim 10$~GeV.

The subject of the present paper is the next-to-minimal supersymmetric
extension of the Standard Model ((M+1)SSM)
\cite{r5,r6,r7,r8,r9,r10,r11,r12,r13} where a gauge singlet superfield $S$ is
added to the Higgs sector. It allows to omit the so-called $\mu$ term $\mu H_1
H_2$ in the superpotential of the MSSM, and to replace it by a Yukawa coupling
(plus a singlet self coupling):

\beq
W = \lambda S H_1 H_2 + \frac{\kappa}{3} \, S^3 + \ldots \label{supot}
\eeq

\noi The superpotential (\ref{supot}) is thus scale invariant, and the
electroweak scale appears only through the susy breaking terms.

In view of ongoing Higgs searches at LEP2 \cite{r14,r16,r17} and, in the near
future, at Tevatron Run II \cite{r15}, it is important to check the model
dependence of bounds on the Higgs mass. In the (M+1)SSM, the upper bound on the
mass $m_1$ of the lightest CP even Higgs\footnote{As there are three CP even
Higgs states in the (M+1)SSM, we denote them as $S_i$ with masses $m_i$,
i=1..3, in increasing order.} differs from the one of the MSSM already at tree
level: now we have \cite{r5,r6}

\beq
m_1^2 \leq M_Z^2 \left ( \cos^2 \! 2\beta + \frac{2\lambda^2}{g_1^2+g_2^2} \,
\sin^2 \! 2\beta \right ) \label{tree}
\eeq

\noi where $g_1$ and $g_2$ denote the $U(1)_Y$ and the $SU(2)_L$ gauge
couplings. Note that, for $\lambda < .53$, $m_1$ is still bounded by $M_Z$ at
tree level. Large values of $\lambda$, $\lambda > .7$, are in any case
prohibited, if one requires the absence of a Landau singularity for $\lambda$
below the GUT scale \cite{r5,r6}.

Loop corrections to $m_1$ have also been considered in the (M+1)SSM \cite{r6}.
Given $m_t$ and assuming again $M_{susy} < 1$~TeV, the upper limit on $m_1$ at
one loop is then $\sim 150$~GeV. Within the constrained (M+1)SSM (the
C(M+1)SSM), where universal soft susy breaking terms at the GUT scale are
assumed \cite{r7,r8,r9,r10}, $\lambda$ is always below $\sim .3$, and the upper
limit on $m_1$ reduces to the one of the MSSM (at one loop) of $\sim 140$~GeV.
Two loop corrections in the (M+1)SSM have recently been considered in
\cite{r13}.

Within the (M+1)SSM this is, however, not the end of the story: It is well
known \cite{r7,r10,r11} that now the lightest Higgs scalar $S_1$ can be
dominantly a gauge singlet state. In this case it decouples from the gauge
bosons and becomes invisible in Higgs production processes, and the lightest
{\em visible} Higgs boson is then actually the second lightest one $S_2$.
Fortunately, under these circumstances $S_2$ cannot be too heavy
\cite{r7,r10,r11}: In the extreme case of a pure singlet lightest Higgs, the
mass $m_2$ of the next-to-lightest Higgs scalar is again below the upper limit
designed originally for $m_1$. In general, however, mixed scenarios can be
realized, with a weakly coupled (but not pure singlet) lightest Higgs, and a
second lightest Higgs above the previous $m_1$ limits. Although analyses of the
Higgs sector including these scenarios in the (M+1)SSM have been presented
before \cite{r11} we find that these should be improved: First, experimental
errors on the top quark pole mass $m_t^{pole} = 173.8 \pm 5.2$~GeV \cite{r18}
have been reduced considerably, leading to stronger constraints on the top
quark Yukawa coupling $h_t$ which determines to a large extent the radiative
corrections to $m_{1,2}$. Second, at least the dominant two loop corrections to
the effective potential should be taken into account, since they are not
necessarily negligible. The purpose of the present paper is thus an analysis of
the allowed masses and couplings to the gauge bosons of the lightest CP even
Higgs scalars in the (M+1)SSM, including present constraints on $m_t$ and a two
loop improvement of the Higgs potential.

In the next section we present our method of obtaining the dominant two loop
terms in the effective potential, and in section~\ref{sec:lhm} we give the
resulting upper bound on the lightest Higgs mass. Albeit this upper limit can
be obtained analytically, the mass of the second lightest Higgs in relation to
the coupling to the gauge bosons requires a numerical analysis. Our methods of
scanning the parameter space of the model in two different scenarios
(constrained and general (M+1)SSM) are presented in section~\ref{sec:par}.
Results on the Higgs masses and couplings, and conclusions are presented in
section~\ref{sec:red}.

\section{Two loop corrections} \label{sec:2lp}

In order to obtain the correct upper limit on the Higgs boson mass in the
presence of soft susy breaking terms, radiative corrections to several terms in
the effective action have to be considered. Let us first introduce a scale $Q
\sim M_{susy}$, where $M_{susy}$ is of the order of the susy breaking terms.
Let us assume that quantum corrections involving momenta $p^2 \gsim Q^2$ have
been evaluated; the resulting effective action $\Gamma_{ef\!f}(Q)$ is then
still of the standard supersymmetric form plus soft susy breaking terms.
Assuming correctly normalized kinetic terms (after appropriate rescaling of the
fields), the $Q$ dependence of the parameters in $\Gamma_{ef\!f}(Q)$ is given
by the supersymmetric $\beta$ functions (valid up to a possible GUT scale
$M_{GUT}$).

Often one is interested in relating the parameters in $\Gamma_{ef\!f}(Q)$ to
more fundamental parameters at $M_{GUT}$. To this end one integrates the
supersymmetric renormalization group equations between $M_{GUT}$ and $Q \sim
M_{susy}$ to one or, if one whishes, to two loop accuracy. Note, however, that
the limits on the Higgs boson mass depend exclusively on the parameters in
$\Gamma_{ef\!f}(Q)$ at the scale $Q \sim M_{susy}$; the two loop contributions
to the effective potential considered below serve to specify this dependence
more precisely. The accuracy to which one has (possibly) related the parameters
at the scale $Q \sim M_{susy}$ to parameters at a scale $M_{GUT}$ is completely
irrelevant for the relation between the Higgs boson mass and the parameters at
the scale  $Q \sim M_{susy}$.

One is left with the computation of quantum corrections to $\Gamma_{ef\!f}$
involving momenta $p^2 \lsim Q^2$. Subsequently the quantum corrections to the
following terms in $\Gamma_{ef\!f}$ will play a role:

\ben
\item[a)] Corrections to the kinetic terms of the Higgs bosons. Due to gauge
invariance the same quantum corrections contribute to the kinetic energy and to
the Higgs-$Z$ boson couplings, which affect the relation between the Higgs vevs
and $M_Z$;
\item[b)] Corrections to the Higgs-top quark Yukawa coupling;
\item[c)] Corrections to the Higgs effective potential. These corrections
could, in principle, be decomposed into contributions to the Yukawa couplings
$\lambda$ and $\kappa$ of eq. (1.2) and the soft terms (these contributions are
the ones proportional to $\ln Q^2$ or, at two loop order, $\ln^2 Q^2$), and
"non-supersymmetric" contributions which are $Q^2$ independent. These latter
contributions to the effective potential are of the orders $(vev)^n$ with $n >
4$ and become small in the case of large soft terms compared to the vevs. Our
results in section 5 are based on the effective potential {\em including} these
contributions (which are not necessarily numerically irrelevant), and there is
no need to perform the decomposition of the radiative corrections to the
effective potential explicitely.
\een

Let us start with the last item: The Higgs effective potential $V_{ef\!f}$ can
be developped in power of $\hbar$ or loops as

\beq
V_{ef\!f} = V^{(0)} + V^{(1)} + V^{(2)} + \ldots .
\eeq

\noi Within the (M+1)SSM, we are interested in the dependence of $V_{ef\!f}$ in
three CP even scalar vevs $h_1$, $h_2$ and $s$ (assuming no CP violation in the
Higgs sector). The tree level potential $V^{(0)}$ is determined by the
superpotential (\ref{supot}) and the standard soft susy breaking terms
\cite{r5,r6,r7,r8,r9,r10,r11}. For completeness, and in order to fix our
conventions, we give here the expression for $V^{(0)}$:

\bea
V^{(0)} & = & m_{H_1}^2 h_1^2 + m_{H_2}^2 h_2^2 + m_{S}^2 s^2 - 2 \lambda
A_\lambda h_1 h_2 s + \frac{2}{3} \, \kappa A_\kappa s^3 \nn \\
& & + \lambda^2 h_1^2 h_2^2 + \lambda^2 (h_1^2 + h_2^2) s^2 - 2 \kappa
\lambda h_1 h_2 s^2 +\kappa^2 s^4 \nn \\
& & + \frac{g_1^2+g_2^2}{8} \, (h_1^2 - h_2^2)^2 . \label{V0}
\eea

The one loop corrections to the effective potential are given by

\beq
V^{(1)} = \frac{1}{64\pi^2} \, \mbox{STr} M^4 \left [ \ln \left (
\frac{M^2}{Q^2} \right ) - \frac{3}{2} \right ] , \label{V1}
\eeq

\noi where we only take top and stop loops into account. The relevant field
dependent masses are the top quark mass

\beq
m_t = h_t h_1 \label{mt}
\eeq

\noi and the stop mass matrix (in the $(T_R^c,T_L)$ basis)

\beq
\left ( \ba{cc}
m_T^2 + m_t^2 & m_t \wt{A}_t \\
m_t \wt{A}_t & m_Q^2 + m_t^2
\ea \right ) , \label{masmat}
\eeq

\noi where $m_T$, $m_Q$ are the stop soft masses and

\beq
\wt{A}_t = A_t - \lambda s \cot\beta \label{defmix}
\eeq

\noi is the so-called stop mixing. In eq.~(\ref{masmat}) we have neglected the
electroweak D terms which would only give small contributions to the effective
potential in the relevant region $m_T,m_Q \gg M_Z$. The masses of the physical
eigenstates $\wt{t}_1,\wt{t}_2$ then read

\beq
m_{\wt{t}_1,\wt{t}_2}^2 = M_{susy}^2 + m_t^2 \pm \sqrt{\delta^2
M_{susy}^4 + m_t^2 \wt{A}_t^2} \label{mstop}
\eeq
\bea
\mbox{with} \quad M_{susy}^2 \equiv \frac{1}{2}(m_Q^2 + m_T^2) \quad
\mbox{and} \quad \delta \equiv \left | \frac{m_Q^2 - m_T^2}{m_Q^2 + m_T^2}
\right | . \label{defsusy}
\eea

\noi Note that the top Yukawa coupling $h_t$ in eq.~(\ref{mt}) and below is
defined at the scale $Q$, cf. the discussion  at the beginning of this section.

In the case of large susy breaking terms compared to the vevs $h_i$, $V^{(1)}$
can be expanded in (even) powers of $h_i$. The terms quadratic in $h_i$ will
not affect the upper bound on the Higgs mass (and can be absorbed into the
unknown soft parameters $m_{H_1}$, $m_{H_2}$ and $A_\lambda$ in (\ref{V0})). In
the approximation where the stop mass splitting $\delta$ is small\footnote{This
approximation is well motivated in the C(M+1)SSM where we take universal soft
terms at the GUT scale. On the other hand, we have checked numerically that, in
the general (M+1)SSM, the lightest Higgs mass takes its maximal value for
$\delta \sim 0$.}, the quartic terms read

\beq
\left . V^{(1)} \right |_{h_i^4} = \frac{3h_t^4}{16\pi^2} \, h_1^4 \left (
\frac{1}{2} \, \wt{X}_t + t \right ) , \label{V1app}
\eeq
\bea
\mbox{where} & & t \equiv \ln \left ( \frac{M_{susy}^2 + m_t^2}{m_t^2} \right
) \label{deft} \\
\mbox{and} & & \wt{X}_t \equiv 2 \,
\frac{\wt{A}_t^2}{M_{susy}^2+m_t^2} \left ( 1 -
\frac{\wt{A}_t^2}{12(M_{susy}^2+m_t^2)} \right ) . \label{Xt}
\eea

\noi In our computations, however, we used the full expression (\ref{V1}) for
$V^{(1)}$; we will use the quartic terms (\ref{V1app}) in the next section only
in order to compare our two loop result to those of refs.~\cite{r4,r13}.

Next, we consider the dominant two loop corrections. These will be numerically
important only for large susy breaking terms compared to $h_i$, hence we will
expand again in powers of $h_i$. Since the terms quadratic in $h_i$ can again
be absorbed into the tree level soft terms, we just consider the quartic terms,
and here only those which are proportional to large couplings: terms $\sim
\alpha_s h_t^4$ and $\sim h_t^6$. Finally, we are only interested in leading
logs (terms quadratic in $t$). The corresponding expression for $V^{(2)}$ can
be obtained from the explicit two loop calculation of $V_{ef\!f}$ in \cite{r3}
or, as we have checked explicitely, from the requirement that the complete
effective potential has to satisfy the renormalization group equations also at
scales $Q < M_{susy}$, provided the non-supersymmetric $\beta$ function for
$h_t$ is used. One obtains in both cases

\beq
V^{(2)}_{LL} = 3 \left ( \frac{h_t^2}{16\pi^2} \right )^2 h_1^4 \left (
32\pi\alpha_s - \frac{3}{2} \, h_t^2 \right ) t^2 . \label{V2}
\eeq

Now, we turn to the quantum corrections to the Higgs boson kinetic terms. They
lead to a wave function renormalization factor $Z_{H_1}$ in front of the $D_\mu H_1
D^\mu H_1$ term with, to order $h_t^2$,

\beq
Z_{H_1} = 1 + 3 \frac{h_t^2}{16\pi^2} \, t \label{zh}
\eeq

Finally, the quantum corrections to the $H_1$-top quark Yukawa coupling $h_t$
have to be considered. After an appropriate rescaling of the $H_1$ and top
quark fields in order to render their kinetic terms properly normalized, these
quantum corrections lead to an effective coupling $h_t(m_t)$ with, to orders
$h_t^2$, $\alpha_s$,

\beq
h_t(m_t) = h_t(Q) \left ( 1 + \frac{1}{32\pi^2} \left ( 32\pi\alpha_s -
\frac{9}{2} \, h_t^2 \right ) t \right ) . \label{htmt}
\eeq

\noi In eqs.~(\ref{zh}) and (\ref{htmt}) the large logarithm $t$ is actually
given by $\ln \left ( \frac{Q^2}{m_t^2} \right )$ where $Q^2$ acts as a UV
cutoff, cf. the discussion at the beginning of this section. In the relevant
region $M_{susy} \gg m_t$ the expression (\ref{deft}) for $t$ can be used here
as well. The (running) top quark mass is then given by

\beq
m_t(m_t) = h_t(m_t) Z_{H_1}^{1/2} h_1 \label{mtrun}
\eeq

\noi and the relation between the pole and running mass, to order $\alpha_s$,
reads

\beq
m_t^{pole} = m_t(m_t) \left ( 1 + \frac{4\alpha_s}{3\pi} \right ) .
\label{mtpole}
\eeq

\section{Upper bound on the lightest Higgs mass} \label{sec:lhm}

In this section we derive an analytic upper bound on the mass of the lightest
Higgs scalar. First, we summarize our contributions to the effective potential.
As it is already known, in the (M+1)SSM the upper bound on the lightest Higgs
mass $m_1$ is saturated when its singlet component vanishes
\cite{r7,r10,r11,r13}. One is then only interested in the $h_i$-dependent part
of the effective potential. Assuming $h_i \ll M_{susy}$, i.e. up to $O(h_i^4)$,
one obtains from eqs.~(\ref{V0}), (\ref{V1app}) and (\ref{V2})

\bea
V_{ef\!f}(h_1,h_2) & = & \wt{m}_1^2 h_1^2 + \wt{m}_2^2 h_2^2 - 
\wt{m}_3^2 h_1 h_2 + \frac{g_1^2+g_2^2}{8} \, (h_1^2-h_2^2)^2 \nn \\
& & + \lambda^2 h_1^2 h_2^2 + \frac{3h_t^2}{16\pi^2} \, h_1^4 \left (
\frac{1}{2} \, \wt{X}_t + t \right ) \nn \\
& & + 3 \left ( \frac{h_t^2}{16\pi^2} \right )^2 h_1^4 \left ( 32\pi\alpha_s -
\frac{3}{2} \, h_t^2 \right ) t^2 \label{above2}
\eea
\noi with
\bea
\wt{m}_1^2 & = & m_{H_1}^2 + \lambda^2s^2 + \mbox{\em rad. corrs.} \quad
, \nn \\
\wt{m}_2^2 & = & m_{H_2}^2 + \lambda^2s^2 + \mbox{\em rad. corrs.} \quad
 , \label{above1} \\
\wt{m}_3^2 & = & 2\lambda s (A_\lambda + \kappa s) + \mbox{\em rad.
corrs.} . \nn
\eea

\noi The radiative corrections in (\ref{above1}) stem from the contributions to
$V^{(1)}$ and $V^{(2)}$ quadratic in $h_i$. In the large $\tan\!\beta$ regime
(which saturates the upper bound on the lightest Higgs in the MSSM), one is
left with only one non-singlet light Higgs $h_1$ and (\ref{above2}) simplifies
to

\beq
V_{ef\!f}(h_1) = \wt{m}_1^2 h_1^2 + \wt{\lambda} h_1^4
\eeq
\noi with
\beq
\wt{\lambda} = \frac{g_1^2+g_2^2}{8} + \frac{3h_t^2}{16\pi^2} \left (
\frac{1}{2} \, \wt{X}_t + t \right ) + 3 \left ( \frac{h_t^2}{16\pi^2}
\right )^2 \left ( 32\pi\alpha_s - \frac{3}{2} \, h_t^2 \right ) t^2 .
\eeq

\noi (Note that in the large $\tan\!\beta$ regime $\wt{A}_t = A_t$ and
no dependence on the (M+1)SSM coupling $\lambda$ is left in
$\wt{\lambda}$.) Now, we can change the variable $h_1$ and replace it by
a variable $h_1'$ in terms of which the kinetic term is properly normalized, so
that we have

\beq
M_Z^2 = \frac{g_1^2+g_2^2}{2} \, h_1'^2 . \label{above4}
\eeq

\noi From eq.~(\ref{zh}) one finds

\beq
h_1^2 \simeq h_1'^2 \left ( 1 - \frac{3h_t^2}{16\pi^2} \, t \right ) .
\label{redef}
\eeq

\noi In terms of $h_1'$ the effective potential reads

\beq
V_{ef\!f}(h_1') = \wt{m}_1'^2 h_1'^2 + \wt{\lambda}' h_1'^4
\label{above3}
\eeq
\noi with
\beq
\wt{m}_1'^2 = \wt{m}_1^2 \left ( 1 - \frac{3h_t^2}{16\pi^2} \,
t \right ) , \quad \wt{\lambda}' = \wt{\lambda} \left ( 1
- \frac{3h_t^2}{16\pi^2} \, t \right )^2 .
\eeq

\noi Second, recall that $h_t$ in the one loop contribution to
eq.~(\ref{above2}) is given by the Yukawa coupling at the scale $Q$. Hence, we
can replace $h_t ( \equiv h_t(Q) )$ in $\wt{\lambda}'$ by $h_t(m_t)$
using eq.~(\ref{htmt}), which allows to relate it directly to the running top
quark mass. Eq.~(\ref{mtrun}) now reads $m_t(m_t) = h_t(m_t) h_1'$.

From (\ref{above3}), one obtains the mass $m_h$ of the lightest non-singlet
Higgs in the case where the singlet decouples (and in the large $\tan\!\beta$
regime)

\beq
m_h^2 = \left . \frac{1}{2} \, \frac{d^2V_{ef\!f}}{dh_1'^{2}}
\right |_{min} = \left . 4 \wt{\lambda}' h_1'^{2} \right |_{min}
.
\eeq

\noi This is just the correct running Higgs mass, but does not include the pole
mass corrections, which involve no large logarithms and which we will neglect
throughout this paper. Using (\ref{above4}) and expanding
$\wt{\lambda}'$ to the appropriate powers of $t$, the expression for
$m_h^2$ becomes\footnote{In eq.~(\ref{above5}) and below in
eq.~(\ref{below1}) we omit the argument of $h_t$ wherever its choice
corresponds to a higher order effect.}

\bea
m_h^2 & = & M_Z^2 \left ( 1 - \frac{3h_t^2}{8\pi^2} \, t \right )
\label{above5} \\
& &  + \frac{3h_t^2(m_t)}{4\pi^2} \, m_t^2(m_t) \left ( \frac{1}{2} \,
\wt{X}_t + t + \frac{1}{16\pi^2} \left ( \frac{3}{2} \, h_t^2 -
32\pi\alpha_s \right ) ( \wt{X}_t + t ) t \right ) \nn
\eea

\noi which agrees with the MSSM result in \cite{r4}. (Note, however, that the
coefficient of the term $\sim \wt{X}_t t$ on the right hand side of
(\ref{above5}) is not necessarily correct, since we would obtain terms of the
same order if we would take into account simple logarithms in the two loop
correction $V^{(2)}$ to the potential.)

The same procedure can be applied for general values of $\tan\!\beta$. Then,
one has to consider the 2x2 mass matrix $\frac{1}{2} ( \partial_{h_i}
\partial_{h_j} V_{ef\!f} ), i,j=1,2$ where the $h_i$ are properly normalized.
Its smallest eigenvalue gives the following upper bound on the mass $m_1$ of
the lightest Higgs boson for arbitrary mixings among the 3 states $(h_1, h_2,
s)$ \cite{r13} (which can be saturated if the lightest Higgs boson has a 
vanishing singlet component)

\bea
m_1^2 \!\!\! & \leq &  \!\!\!\! M_Z^2 \left ( \cos^2 \! 2\beta +
\frac{2\lambda^2}{g_1^2+g_2^2} \, \sin^2 \! 2\beta \right ) \left ( 1 -
\frac{3h_t^2}{8\pi^2} \, t \right ) \label{below1} \\
& & \!\!\!\! + \frac{3h_t^2(m_t)}{4\pi^2} \, m_t^2(m_t) \sin^2 \! \beta \left (
\frac{1}{2} \, \wt{X}_t + t + \frac{1}{16\pi^2} \left ( \frac{3}{2} \,
h_t^2 - 32\pi\alpha_s \right ) ( \wt{X}_t + t ) t \right ) . \nn
\eea

\noi The only difference between the MSSM bound \cite{r4} and (\ref{below1}) is
the 'tree level' term $\sim \lambda^2 \sin^2 \! 2\beta$. This term is important
for moderate values of $\tan\!\beta$. Hence, the maximum of the lightest Higgs
mass in the (M+1)SSM is not obtained for large $\tan\!\beta$ as in the MSSM,
but rather for moderate $\tan\!\beta$ (as confirmed by our numerical analysis,
cf. section~\ref{sec:red}). On the other hand, the radiative corrections are
identical in the (M+1)SSM and in the MSSM. In particular, the linear dependence
in $\wt{X}_t$ is the same in both models. Hence, from eq.~(\ref{Xt}), the upper
bound on $m_1^2$ is maximized for $\wt{X}_t=6$ (corresponding to
$\wt{A}_t=\sqrt{6} M_{susy}$, the 'maximal mixing' case), and minimized for
$\wt{X}_t=0$ (corresponding to $\wt{A}_t=0$, the 'no mixing' case).

\section{Parametrization of the (M+1)SSM} \label{sec:par}

Eq.~(\ref{below1}) gives an upper bound on the lightest Higgs mass $m_1$ {\em
regardless of its coupling to the gauge bosons}. In the extreme case of a pure
singlet lightest Higgs, the next-to-lightest Higgs is non-singlet and the upper
bound (\ref{below1}) actually applies to $m_2$. On the other hand, it can occur
that the lightest Higgs is weakly coupled to gauge bosons (without being a pure
singlet) and $m_2$ is above the limit (\ref{below1}). This case requires a
numerical analysis, which will be performed in the next section. First, let us
present our methods of scanning the parameter space of the (M+1)SSM.

Not counting the known gauge couplings, the parameters of the model are

\beq
\lambda \,,\; \kappa \,,\; h_t \,,\; A_\lambda \,,\; A_\kappa \,,\; A_t \,,\;
m_{H_1}^2 \,,\; m_{H_2}^2 \,,\; m_S^2 \,,\; m_Q^2 \,,\; m_T^2 \label{paraEW}
\eeq

\noi where $h_t$ is eventually fixed by the top mass and an overall scale of
the dimensionful parameters by the $Z$ mass. Now, let us see how to handle this
high dimensional parameter space in two different scenarios.

\subsection{Constrained (M+1)SSM}

In the C(M+1)SSM the soft terms are assumed universal at the GUT scale, the
global minimum of the effective potential has to be the global minimum and
present experimental constraints on the sparticle and Higgs masses are applied.
The free parameters can be chosen as the GUT scale dimensionless parameters

\beq
\lambda_0 \,,\; \kappa_0 \,,\; h_{t0} \,,\; \frac{A_0}{M_{1/2}} \,,\;
\frac{m_0^2}{M_{1/2}^2} \label{paraGUT}
\eeq

\noi where $A_0$, $M_{1/2}$ and $m_0^2$ are the universal trilinar coupling,
gaugino mass and scalar mass respectively. In order to scan the 5-dimensional
parameter space of the C(M+1)SSM, we proceed as in refs.~\cite{r7,r9}: 

First, we scan over the GUT scale parameters (\ref{paraGUT}) and integrate
numerically the renormalization group equations \cite{r5} down to the susy
scale in each case.

Then, we minimize the complete two loop effective potential in order to obtain
the Higgs vevs $h_1,h_2,s$. In principle, we could have followed the same
procedure as in section~\ref{sec:lhm} to obtain the dominant two loop
corrections, i.e. replacing $h_1$ by $h_1'$ and $h_t$ by $h_t(m_t)$.
However, in order to obtain numerically correct results also in the regime
$M_{susy} < 1$~TeV, we did not expand $V^{(1)}$ in powers of $h_i/M_{susy}$,
i.e. we used the full expression (\ref{V1}) for $V^{(1)}$. Then it becomes
inconvenient to perform the field redefinition (\ref{redef}), which is
implicitly non-linear due to the $h_1$ dependence of $t$ via $m_t$.
Therefore we proceed differently: For a given set of low energy parameters,
which are implicitly obtained at the scale $Q \sim M_{susy}$, we minimize
directly

\beq
V_{ef\!f} = V^{(0)} + V^{(1)} + V^{(2)} \label{Veff}
\eeq

\noi with $V^{(0)}$ as in (\ref{V0}), $V^{(1)}$ as in (\ref{V1}) and $V^{(2)}$
as in (\ref{V2}). Points in the parameter space leading to deeper unphysical
minima of the effective potential with $h_i=0$ or $s=0$ are removed.

The overall scale is then fixed by relating the vevs $h_i$ to the physical $Z$
mass through

\beq
M_Z^2 = \frac{1}{2} \, (g_1^2 + g_2^2) (Z_{H_1}^2h_1^2 + h_2^2) \label{Zmass}
\eeq

\noi with $Z_{H_1}$ as in  eq.~(\ref{zh}). Next, we throw away all points in
the parameter space where the top quark mass (including corrections
(\ref{htmt}) to $h_t$) does not correspond to the measured $m_t^{pole} = 173.8
\pm 5.2$~GeV. We also ask for sfermions with masses $m_{\wt f} \gsim
M_Z/2$ and gluinos with masses $m_{\wt g} \gsim 200$~GeV.

Finally, the correct 3x3 Higgs mass matrix is related to the matrix of second 
derivatives of the Higgs potential at the minimum after dividing $\frac{1}{2}
\partial^2_{h_1} V_{ef\!f}$ by $Z_{H_1}$, and $\frac{1}{2} \partial_{h_1}
\partial_{h_2} V_{ef\!f}$ and $\frac{1}{2} \partial_{h_1} \partial_s V_{ef\!f}$
by $Z_{H_1}^{1/2}$. For each point in the parameter space, we then obtain the
two loop Higgs boson masses and couplings to gauge bosons. Then, we apply
present constraints from negative Higgs search at LEP (cf.
section~\ref{sec:red} for details).

The results in section~5 are based on scannings over $\sim 10^6$ points in the
parameter space. The essential effect of all constraints within the C(M+1)SSM
is to further reduce the allowed range for the Yukawa coupling $\lambda$ to
$\lambda \lsim .3$.

\subsection{General (M+1)SSM}

In the general (M+1)SSM, we only assume that we are in a {\em local} minimum of
the effective potential (\ref{Veff}) and the running Yukawa couplings $\lambda
, \kappa , h_t$ are free of Landau singularities below the GUT scale. In order
to scan the high dimensional parameter space (\ref{paraEW}) of the general
(M+1)SSM we proceed as follows:

First, we use the three minimization equations of the full effective potential
(\ref{Veff}) with respect to $h_1$, $h_2$ and $s$ in order to eliminate the
parameters $m_{H_1}^2$, $m_{H_2}^2$ and $m_S^2$ in favour of the three Higgs
vevs. Using the relation (\ref{Zmass}), we replace $h_1, h_2$ by $\tan\!\beta$
and $M_Z$. Finally, eqs.~(\ref{htmt}), (\ref{mtrun}) and (\ref{mtpole}) allow
us to express $h_t$ in terms of $m_t^{pole}$ and the other parameters.

We are then left with six 'tree level' parameters $\lambda$, $\kappa$,
$A_\lambda$, $A_\kappa$, $s$, $\tan\!\beta$, and three parameters appearing
only through the radiative corrections, which we choose as $\wt{A}_t$,
$M_{susy}$ and $\delta$, as defined in eqs.~(\ref{defmix}) and (\ref{defsusy}).

Requiring that the Yukawa couplings are free of Landau singularities below the
GUT scale and using the renormalization group equations of the (M+1)SSM
\cite{r5}, one obtains upper limits on $\lambda , \kappa , h_t$ at the susy
scale. The latter turns into a lower bound on $\tan\!\beta$ depending mainly on
$m_t^{pole}$ and $M_{susy}$. As expected from eq.~(\ref{below1}), we observe
that upper limits on Higgs masses are obtained when $\lambda$ is maximal.  From
the renormalization group equations, one finds that the upper limit on
$\lambda$ increases with decreasing $\kappa$, thus we choose $\kappa \sim 0$
and $\lambda = \lambda_{max} \sim .7$ (which still depends on $h_t$, i.e. on
$\tan\!\beta$).

As already mentionned, one can see from eq.~(\ref{below1}) that the lightest
Higgs mass is maximized for moderate values of $\tan\!\beta$. Hence, except in
fig.~4 where $\tan\!\beta$ varies, we fix $\tan\!\beta = 2.7$ which, as we
shall see, maximizes the Higgs masses for $m_t^{pole} = 173.8$~GeV.

Unless stated otherwise, the upper limits on the Higgs masses presented in the
next section are given in the maximal mixing scenario ($\wt{A}_t =
\sqrt{6} M_{susy}$). We have also found that Higgs masses are maximized for
small values of $\delta$ and fixed $\delta = 0$ (thus $m_Q = m_T = M_{susy}$).
In order to obtain the results presented in the next section, we have used
numerical routines to maximize the Higgs masses with respect to the remaining
three parameters $A_\lambda, A_\kappa, s$.

\section{Reduced couplings versus mass bounds} \label{sec:red}

Let us start with the mass $m_1$ of the lightest Higgs scalar,
independently of its coupling to gauge bosons. The upper limit on $m_1$
in the general (M+1)SSM is plotted in fig.~1 (straight line) as a
function of $M_{susy}$ (for $m_t^{pole} = 173.8$~GeV). This limit is
well above the one of the MSSM because of the additional tree level
contribution to $m_1^2$ proportional to $\lambda^2 M_Z^2$ (cf.
eq.~(\ref{tree})). At $M_{susy} = 1$~TeV we have $m_1 \leq 133.5$~GeV
(in agreement with the analytic approximation (\ref{below1})); at
$M_{susy} = 3$~TeV this upper limit increases only by $\sim 3$~GeV.
This weak dependence on $M_{susy}$ is due to the negative two loop
contributions to $m_1$.

Within the C(M+1)SSM, the combined constraints on the parameter space require
$\lambda$ to be small, $\lambda \lsim .3$ \cite{r7,r9}. Accordingly, the upper
limit on $m_1$ is very close to the one of the MSSM. It is shown as crosses in
fig.~1, and reaches 120~GeV at $M_{susy} = 1$~TeV. In the following, we shall
assume $M_{susy} = 1$~TeV.

A sideremark on the behavior for small $M_{susy}$ is in order: From
eq.~(\ref{mstop}), it is obvious that, in the assumed limit $\delta \rightarrow
0$, the assumption of maximal stop mixing ($\wt{A}_t = \sqrt{6}
_{susy}$) cannot be maintained for

\beq
\frac{\sqrt{6} - \sqrt{2}}{2} \, m_t < M_{susy} < \frac{\sqrt{6} + \sqrt{2}}{2}
\, m_t ,
\eeq

\noi because it would imply a negative stop mass squared. Therefore, in the
general (M+1)SSM, we choose $\wt{A}_t$ in this regime such that the
lightest stop mass squared remains positive. On the other hand, within the
C(M+1)SSM, where soft susy breaking terms are related, the limit $M_{susy}$
small is not feasable since it would contradict the negative results on
sparticle searches.

As discussed in the introduction, the upper limit on $m_1$ is not necessarily
physically relevant, since the coupling of the lightest Higgs to the $Z$ boson
can be very small. Actually, this phenomenon can also appear in the MSSM, if
$\sin^2(\beta-\alpha)$ is small. However, the CP odd Higgs boson $A$ is then
necessarily light ($m_A \sim m_h < M_Z$ at tree level), and the process $Z
\rightarrow h A$ can be used to cover this region of the parameter space in the
MSSM. In the (M+1)SSM, a small gauge boson coupling of the lightest Higgs $S_1$
is usually related to a large gauge singlet component, in which case no
(strongly coupled) light CP odd Higgs boson is available. Hence, Higgs searches
in the (M+1)SSM have possibly to rely on the search for the second lightest
Higgs scalar $S_2$.

Let us now define $R_i$ as the square of the coupling $ZZS_i$ divided by the
corresponding standard model Higgs coupling:

\beq
R_i = (S_{i1}\sin\!\beta + S_{i2}\cos\!\beta)^2
\eeq

\noi where $S_{i1}, S_{i2}$ are the $H_1, H_2$ components of the CP even Higgs
boson $S_i$, respectively. Evidently, we have $0 \leq R_i \leq 1$ and unitarity
implies

\beq
\sum_{i=1}^3 R_i = 1 . \label{Rsum}
\eeq

Fortunately, as it was already mentionned, in the extreme case $R_1 \rightarrow
0$ the upper limit on $m_2$ is the same as the above upper limit on $m_1$. On
the other hand, scenarios with, e.g., $R_1 \sim R_2 \sim 1/2$ are possible. In
the following we will discuss these situations in detail.

We are interested in upper limits on the two lightest CP even Higgs bosons
$S_{1,2}$. These are obtained in the limit where the third Higgs, $S_3$, is
heavy and decouples, i.e. $R_3 \sim 0$ (This is the equivalent of the so called
decoupling limit in the MSSM: the upper bound on the lightest Higgs $h$ is
saturated when the second Higgs $H$ is heavy and decouples). Hence, we have
$R_1 + R_2 \simeq 1$.

In the regime $R_1 \geq 1/2$ experiments will evidently first discover the
lightest Higgs (with $m_1 \leq 133.5$~GeV for $M_{susy} = 1$~TeV). The 'worst
case scenario' in this regime corresponds to $m_1 \simeq 133.5$~GeV, $R_1 \simeq
1/2$; the presence of a Higgs boson with these properties has to be excluded in
order to test this part of the parameter space of the general (M+1)SSM.

The regime where $R_1 < 1/2$ (and hence $1/2 < R_2  \leq 1$) is more delicate:
here the lightest Higgs may escape detection because of its small coupling, and
it may be easier to detect the second lightest Higgs. In fig.~2 we show the
upper limit on $m_2$ as a function of $R_2$ in the general (M+1)SSM as a thin
straight line. For $R_2 \rightarrow 1$ (corresponding to $R_1 \rightarrow 0$)
we obtain the announced result: the upper limit on a Higgs boson with $R
\rightarrow 1$ is always given by the previous upper limit on $m_1$, even if
the corresponding Higgs boson is actually the second lightest one. The same
applies, of course, to the C(M+1)SSM where the upper limit on $m_2$ is also
indicated as crosses in fig.~2. In the following we will discuss this
'delicate' regime, $R_1 < 1/2$ and $1/2 < R_2  \leq 1$, in some detail:

Fortunately, one finds that the upper limit on $m_2$ is saturated only when the
mass $m_1$ of the lightest Higgs boson tends to 0.  Clearly, one has to take
into account the constraints from Higgs boson searches which apply to reduced
couplings $R < 1/2$ -- i.e. lower limits on $m_1$ as a function of $R_1 \simeq
1 - R_2$ -- in order to obtain realistic upper limits on $m_2$ vs $R_2$.

Lower limits on $m_1$ as a function of $R_1$ (in the regime $R_1 < 1/2$) have
been obtained at LEP \cite{r16}. We use the following analytic approximation
for the constraints on $R_1$ vs $m_1$ in this regime:

\beq
\log_{10} \! R_1 < \frac{m_1}{45 \mbox{GeV}} - 2 \label{lep}
\eeq

\noi The resulting upper limit on $m_2$ is shown in fig.~2 as a thick straight
line. This constraint is automatically included in the C(M+1)SSM results
(crosses). Present and future Higgs searches at LEP will lead to more stringent
constraints in the regime $.1 < R_1 < 1/2$ \cite{r17}. We approximate the
possible constraints from a run at 198~GeV c.m. energy and 200~pb$^{-1}$ by

\beq
\ln \! R_1 < 2 \left ( \frac{m_1}{98 \mbox{GeV}} \right )^4 - 3 \label{lep2}
\eeq

\noi The resulting upper limit on $m_2$ is shown in fig.~2 as a thick dashed
line.

It would be desirable to have the upper limit on $m_2$ in the general (M+1)SSM
for arbitrary lower limits on $m_1$ as a function of $R_1$. To this end we have
produced fig.~3. The different dotted curves show the upper limit on $m_2$ as a
function of $R_2$ for different lower limits on $m_1$ (as indicated on each
curve) as a function of $R_1$ (as indicated at the top of fig.~3).

In practice, fig.~3 can be used to obtain upper limits on the mass $m_2$, in
the regime $R_1 < 1/2$, for arbitrary experimental lower limits on the mass
$m_1$: For each value of the coupling $R_1$, which corresponds to a vertical
line in fig.~3, one has to find the point where this vertical line crosses the
dotted curve associated to the corresponding experimental lower limit on $m_1$.
Joining these points by a curve leads to the upper limit on $m_2$ as a function
of $R_2$. We have indicated again the present LEP limit (\ref{lep}), already
shown in fig.~2, which excludes the shaded region ($m_2 > 172.5$~GeV for $R_2 =
.5$, $m_2 > 150$~GeV for $R_2 = .75$, {\it etc}). We have also shown again the
possible LEP2 constraints on $m_2$ arising from (\ref{lep2}) as a thick dashed
line.

Lower experimental limits on a Higgs boson with $R > 1/2$  restrict the allowed
regime for $m_2$ (for $R_2 > 1/2$) in fig.~3 from below. The present lower
limits on $m_2$ from LEP are not visible in fig.~3, since we have only shown
the range $m_2 > 130$~GeV. Possibly Higgs searches at the Run II of the
Tevatron push the lower limits on $m_2$ upwards into this range. This would be
necessary if one aims at an exclusion of the 'delicate' regime of the (M+1)SSM:
Then, lower limits on the mass $m_2$ -- for any value of $R_2$ between $1/2$
and 1 -- of at least 133.5~GeV are required; the precise experimental lower
limits on $m_2$ as a function of $R_2$, which would be needed to this end, will
depend on the achieved lower limits on $m_1$ as a function of $R_1$ in the
regime $R_1 < 1/2$.

In principle, from eq. (\ref{Rsum}), one could have $R_2 > R_1$ with $R_2$ as
small as $1/3$. However, in the regime $1/3 < R_2 < 1/2$, the upper bound on
$m_2$ as a function of $R_2$ for different fixed values of $m_1$ can only be
saturated if $R_1 = R_2$. Then it is sufficient to look for a Higgs boson with
a coupling $1/3 < R < 1/2$ and a mass $m \lsim 133.5$~GeV to cover this region
of the parameter space of the (M+1)SSM.

Finally, we consider the dependence of the upper bounds on $m_{1,2}$ on
$\tan\!\beta$ and the top quark pole mass. In fig.~4 we plot the upper limit on
$m_{1,2}$ (for $R_{1,2} = 1$) against $\tan\!\beta$ for $m_t^{pole} =
173.8$~GeV as a thick straight line. Remarkably, as announced before, this
$\tan\!\beta$ dependence is very different from the MSSM: the maximum is
assumed for $\tan\!\beta \simeq 2.7$  (with $m_{1,2} \simeq 133.5$~GeV in
agreement with figs.~2,3). The origin of this $\tan\!\beta$ dependence is the
tree level contribution $\sim \lambda^2\sin^2\!\beta$ to (3.11). The height and
the location of the maximum varies somewhat with $m_t^{pole}$; the thick dashed
and dotted curves correspond to $m_t^{pole} = 173.8 \pm 5.2$~GeV, respectively.
The absolute maximum is at $\tan\!\beta \simeq 3$ with $m_{1,2} \simeq
135$~GeV.

In the 'delicate' regime, where one has to search for the second lightest Higgs
with $R_2$ between $1/2$ and $1$, one could worry whether the $\tan\!\beta$
dependence of the upper limit on $m_{2}$ is different. This is not the case: As
a thin straight line we show the upper limit on $m_2$ in the extreme case $R_2
= 1/2$ and $m_t^{pole} = 173.8$~GeV (where the LEP constraint (\ref{lep}) is
taken into account), which assumes again its maximum for $\tan\!\beta \simeq
2.7$ (now with $m_{2} \simeq 172.5$~GeV in agreement with figs.~2,3). As
above, the thin dashed and dotted curves correspond to $m_t^{pole} = 173.8 \pm
5.2$~GeV, respectively, and the absolute maximum is at $\tan\!\beta \simeq 3$
with $m_2 \simeq 175.5$~GeV. Within the C(M+1)SSM, where $\lambda$ is small,
the dependence of the upper limit on $m_2$ on $\tan\!\beta$ ressembles more to
the one of the MSSM as shown as crosses in fig.~4.

To conclude, we have studied the CP even Higgs sector of the general (M+1)SSM
and the C(M+1)SSM including the dominant two loop corrections to the effective
potential. We have emphasized the need to search for Higgs bosons with reduced
couplings, which are possible within this model. Our main results are presented
in fig.~3, which allows to obtain the constraints on the Higgs sector of the
model both from searches for Higgs bosons with weak coupling ($R < 1/2$), and
strong coupling ($R > 1/2$). The necessary (but not sufficient) condition for
testing the complete parameter space of the (M+1)SSM is to rule out a CP even
Higgs boson with a coupling $1/3 < R < 1$ and a mass below 135~GeV. The
sufficient condition (i.e. the precise upper bound on $m_2$ vs $R_2$) depends
on the achieved lower bound on the mass of a 'weakly' coupled Higgs (with $0 <
R < 1/2$) and can be obtained from fig.~3. At the Tevatron this would probably
require an integrated luminosity of up to 30 fb$^{-1}$ \cite{r15}. If this
cannot be achieved, and no Higgs is discovered, we will have to wait for the
results of the LHC in order to see whether supersymmetry beyond the MSSM is
realized in nature.

\newpage\section*{Figure Captions}
\newcounter{fig}
\begin{list}{\bf Figure \arabic{fig}:}{\usecounter{fig}}

\item Upper limits on the mass $m_1$ of the lightest CP even Higgs boson versus
$M_{susy}$ in the general (M+1)SSM (straight line) and the C(M+1)SSM (crosses).
\label{fig1}

\item Upper limits on the mass $m_2$ of the second lightest CP even Higgs (in
the regime $R_2 > 1/2$) against $R_2$ in the general (M+1)SSM (thin straight
line); the general (M+1)SSM with LEP constraints (\ref{lep}) (thick straight
line); the general (M+1)SSM with expected LEP2 constraints (\ref{lep2}) (thick
dashed line); the C(M+1)SSM with LEP constraints (\ref{lep}) (crosses).
\label{fig2}

\item Upper limits on the mass $m_2$ against $R_2$, for different lower limits
on the mass $m_1$ (as indicated on each line in GeV) of the lightest Higgs
boson for $1/2 < R_2 < 1$. $R_1=1-R_2$ is shown on the the top axis. The
boundary of the shaded area corresponds to the thick line in fig.~\ref{fig2},
also the dashed line is the same as in fig.~\ref{fig2}. \label{fig3}

\item Upper limits on $m_{1,2}$ with $R_{1,2} = 1$ (thick lines), and upper
limits on $m_2$ with $R_2 = 1/2$ (thin lines) versus $\tan\beta$ in the general
(M+1)SSM for $m_t^{pole} = 173.8$~GeV (straight), $179$~GeV (dashed) and
$168.6$~GeV (dotted); upper limit on $m_2$ in the C(M+1)SSM (crosses) for
$m_t^{pole} = 173.8 \pm 5.2$~GeV. The LEP constraints (\ref{lep}) are taken
into account in each case. \label{fig4}

\end{list}

\begin{figure}[p]
\unitlength1cm
\begin{picture}(20,20)
\put(-3.3,-4.5){\epsfig{file=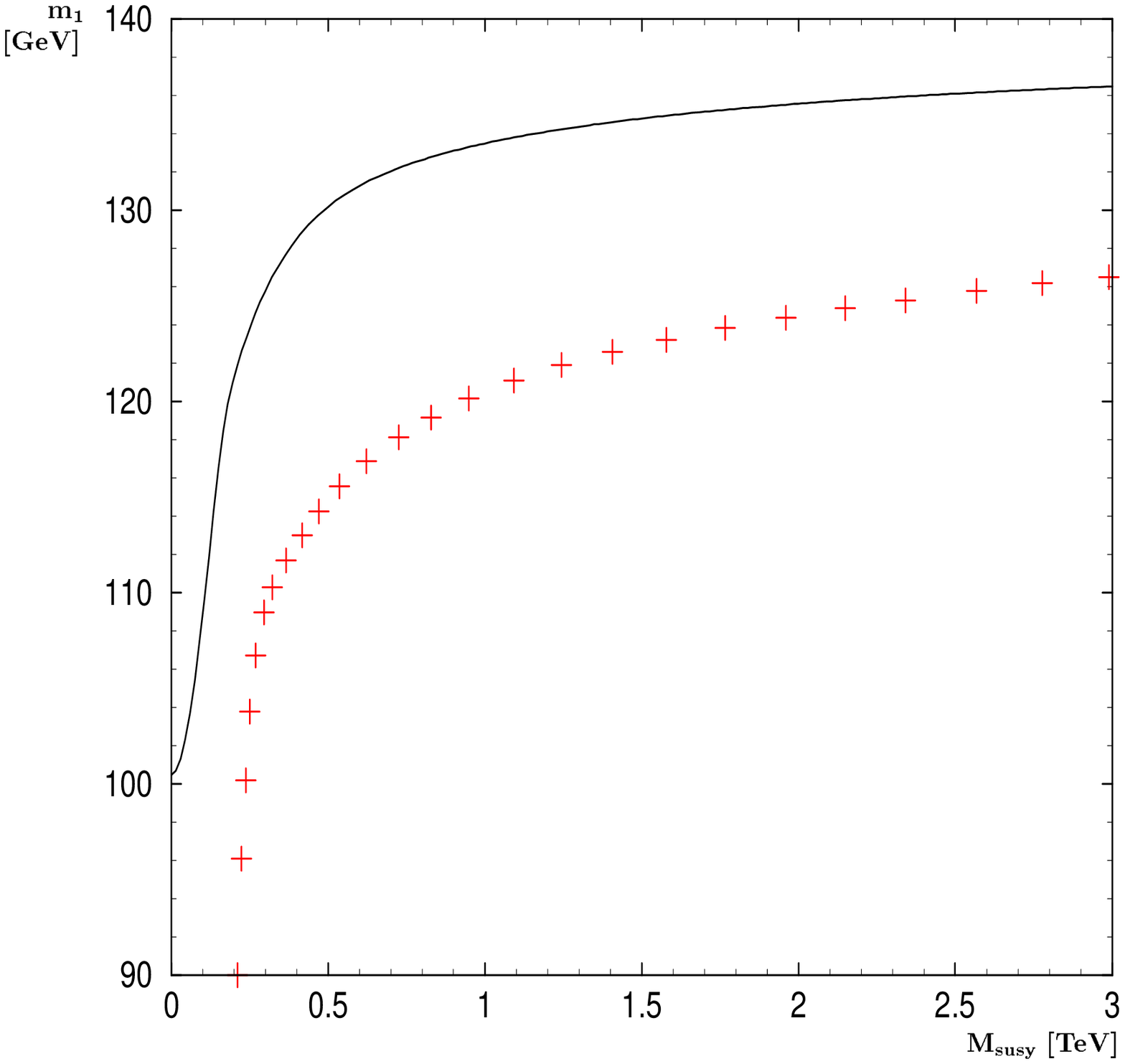}}
\put(6,0){\bf Figure 1}
\end{picture}
\end{figure}

\begin{figure}[p]
\unitlength1cm
\begin{picture}(20,20)
\put(-3.3,-4.5){\epsfig{file=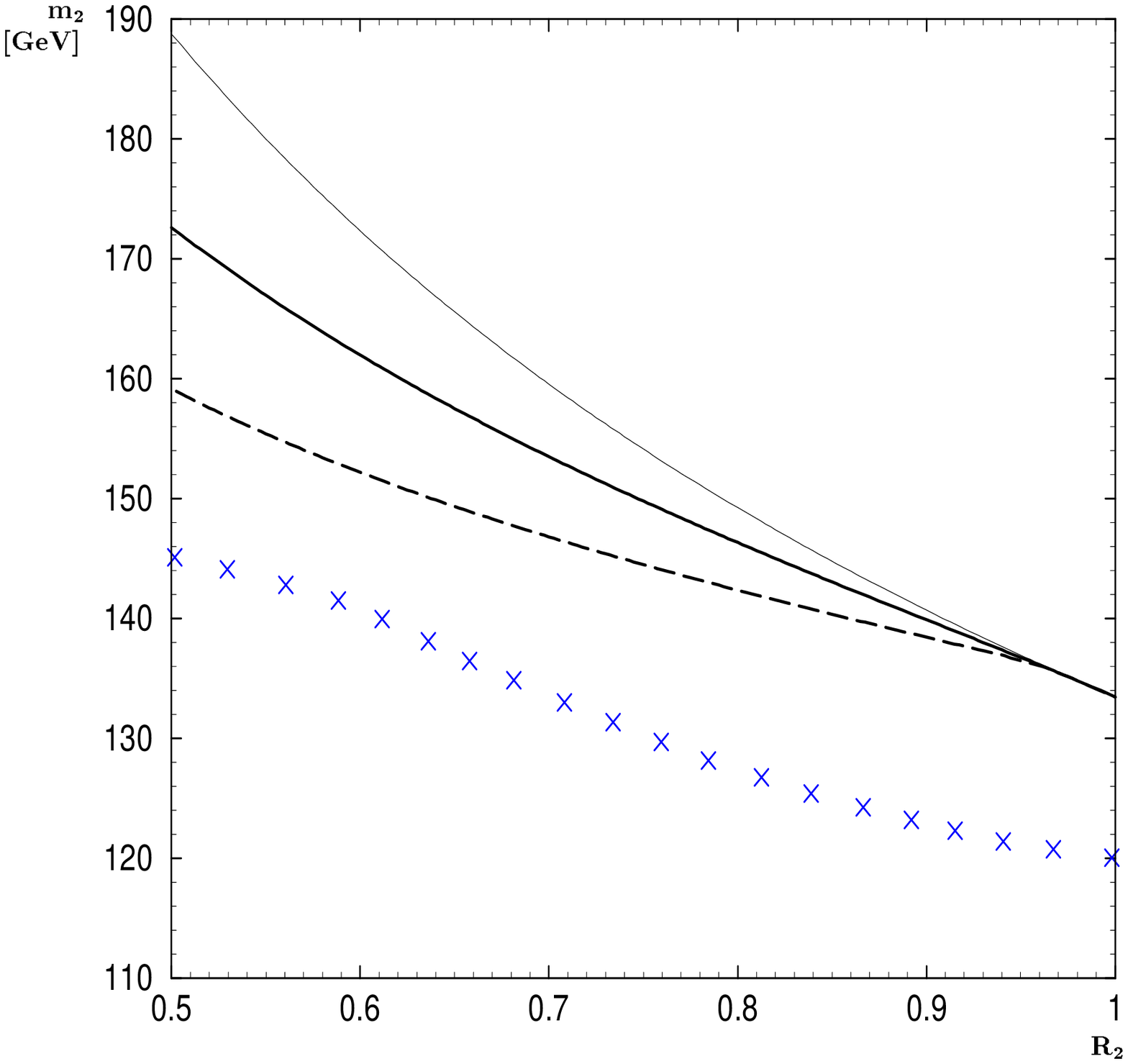}}
\put(6,0){\bf Figure 2}
\end{picture}
\end{figure}

\begin{figure}[p]
\unitlength1cm
\begin{picture}(20,20)
\put(-3.3,-4.5){\epsfig{file=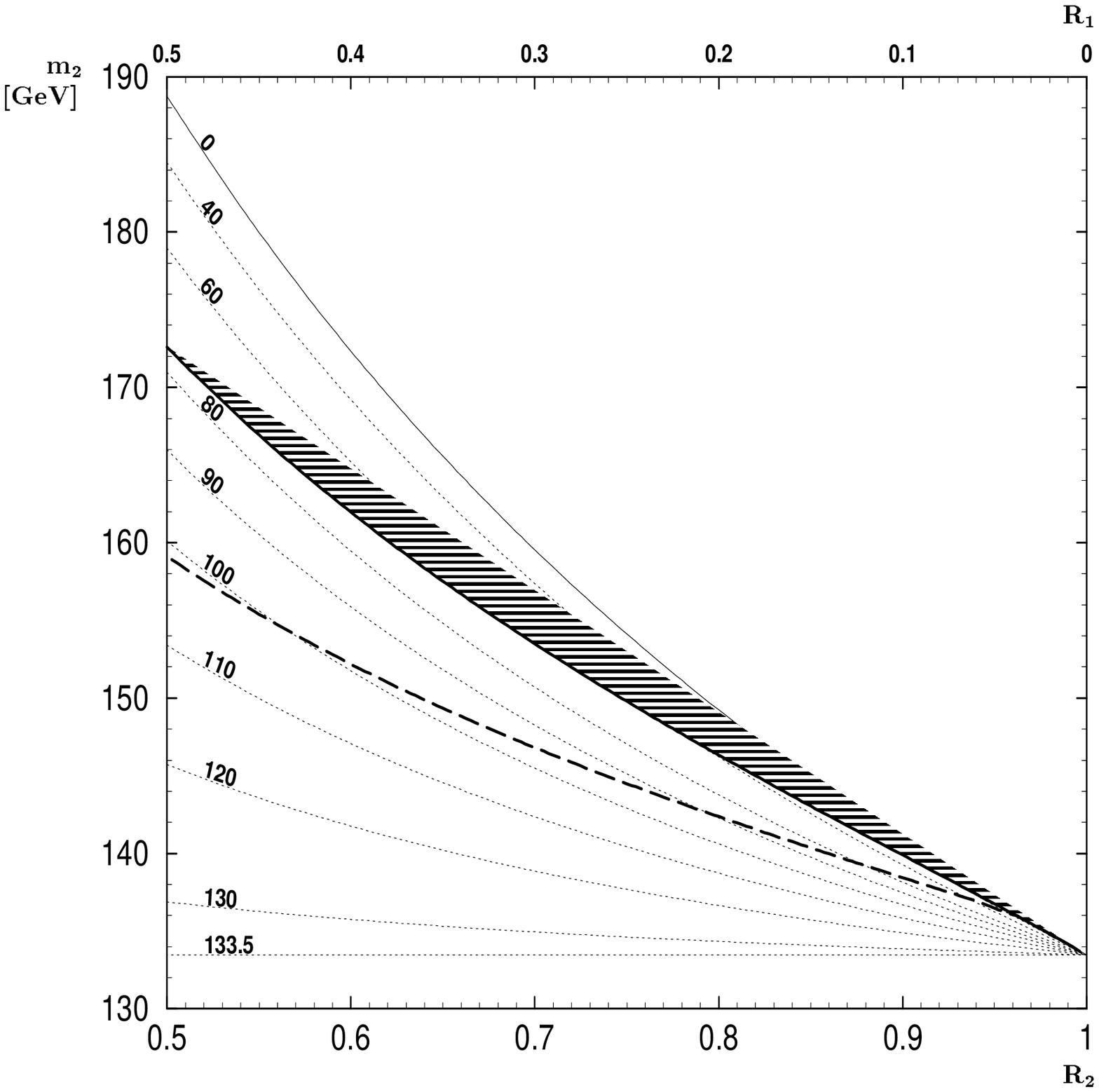}}
\put(6,0){\bf Figure 3}
\end{picture}
\end{figure}

\begin{figure}[p]
\unitlength1cm
\begin{picture}(20,20)
\put(-3.3,-4.5){\epsfig{file=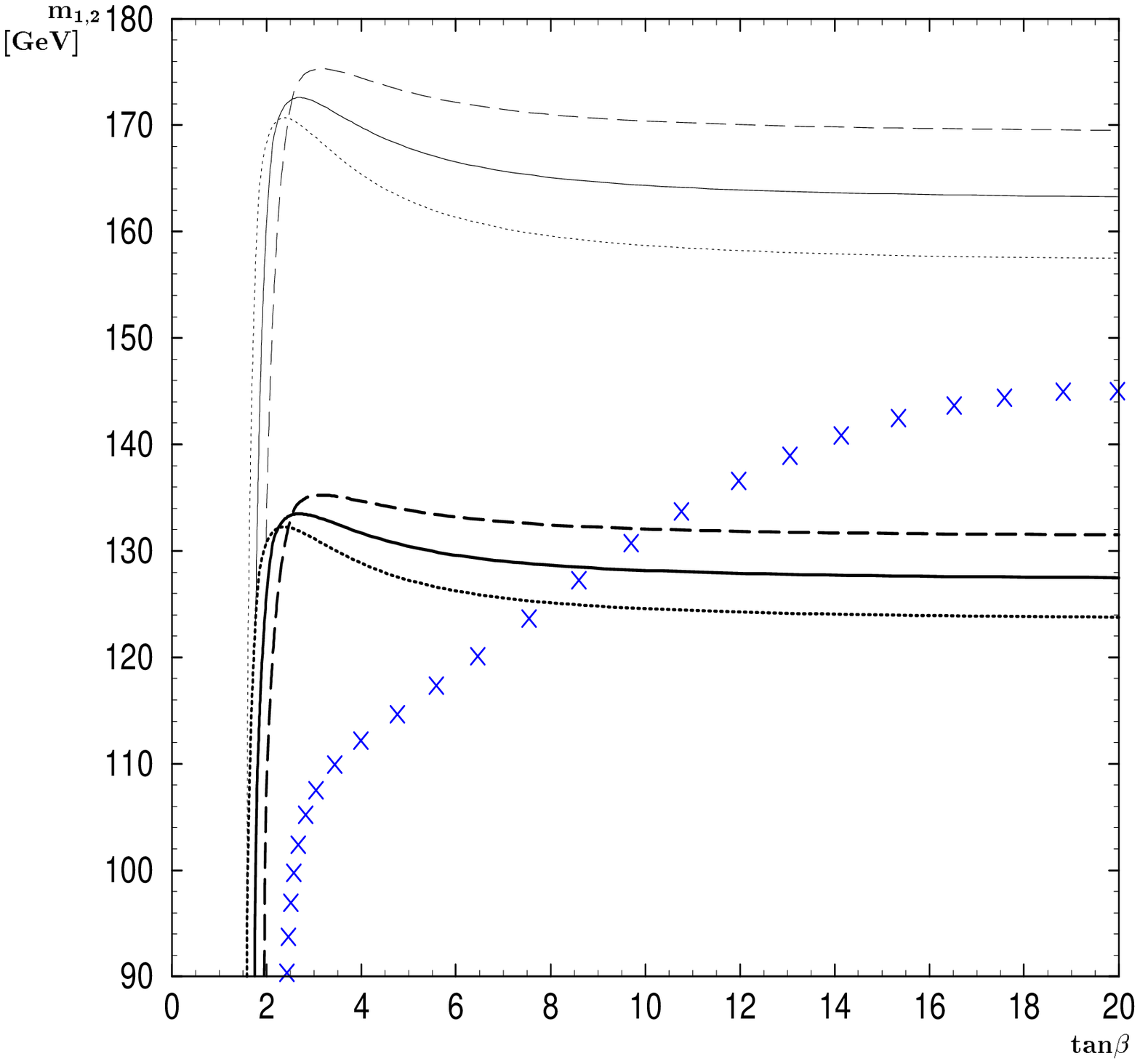}}
\put(6,0){\bf Figure 4}
\end{picture}
\end{figure}

\end{document}